\documentclass[11pt]{amsart}
\usepackage{amscd}
\usepackage{amsmath}
\usepackage{graphicx}
\usepackage{amsfonts}
\usepackage{amssymb}
\begin{document}

%20120319 21:22

\title[Measurement-induced nonlocality over two-sided projective measurements]
{Measurement-induced nonlocality over two-sided projective measurements}

%\date{\today}

\author{Yu Guo}
%%\address[Y. Guo]
%{Department of Mathematics, Shanxi Datong University, Datong 037009,
%China; \\
%Institute of Optoelectroincs Engineering, Department of Physics and
%Optoelectroincs, Taiyuan University of Technology, Taiyuan 030024,
%China}
%\email{guoyu3@yahoo.com.cn}

%\author{Jinchuan Hou}
%\address[J. Hou]{Department of
%Mathematics,
%Taiyuan University of Technology,
% Taiyuan 030024,
%  P. R. China}
%\email{jinchuanhou@yahoo.com.cn, houjinchuan@tyut.edu.cn}


%\thanks{{\it 2010 Mathematics Subject Classification.} Primary
%47B47; 47A63; 47C15.}
%\thanks{This work is partially supported by  Research Fund for the Doctoral Program of Higher Education of China (20101402110012),
% Tianyuan Funds of China (11026161) and
%Foundation of Shanxi University.}

\thanks{Department of Mathematics, Shanxi Datong University, Datong 037009,
China; Institute of Optoelectroincs Engineering, Department of Physics and
Optoelectroincs, Taiyuan University of Technology, Taiyuan 030024,
China}
\thanks{Email: guoyu3@yahoo.com.cn}

\maketitle

\begin{abstract}
Measurement-induced nonlocality (MiN), introduced by Luo and Fu [Phys. Rev. Lett. 106(2011)120401], is a kind of quantum correlation that beyond entanglement and even beyond quantum discord. Recently, we extended MiN to infinite-dimensional bipartite system [arXiv:1107.0355].
MiN is defined over one-sided projective measurements. In this letter we introduce a measurement-induced nonlocality over two-sided projective
measurements. The nullity of this two-sided MiN is characterized, a formula for calculating two-sided MiN for pure states is proposed, and a lower bound of (two-sided) MiN for maximally entangled mixed states is given.
In addition, we find that (two-sided) MiN is not continuous.
The two-sided geometric measure of quantum discord (GMQD) is introduced in [Phys. Lett. A 376(2012)320--324]. We extend it to infinite-dimensional system and then compare it with the two-sided MiN. Both finite- and infinite-dimensional cases are considered.
\end{abstract}

%03.67.Mn, 03.65.Db, 03.65.Ud.

\section{Introduction}

Quantum correlation in composite quantum states lies at the heart of quantum world \cite{Nielsen,Horodecki1,Guhne,Modi,Lewenstein,Chen,
Rudolph,Guo1,Horodecki3,Hou2,Hou3,Hou4,Hou5,Hou6,Hou7,Ollivier,Luo1,Luoshun,Hou}.
Recently, much interest has been devoted to the study of quantum correlations that may arise without
entanglement, such as quantum discord (QD) \cite{Ollivier}, measurement-induced nonlocality (MiN)\cite{Luo1} and quantum deficit \cite{Oppenheim}, etc. These quantum correlations
are argued to be responsible for certain quantum computation and quantum information tasks \cite{Modi,Luo1,Dakic,Qasimi,Roa}.

MiN is one of the ways to detect quantum correlation by locally
invariant projective measurements \cite{Luo1}. Locally invariant measurements can not affect global states in classical theory but this is possible in quantum theory. So MiN is a type of quantum correlation which can only exists in quantum domain.
It is indicated in \cite{Luo1} that MiN may
be applied in quantum cryptography, general quantum dense
coding \cite{Mattle,Li}, remote state control \cite{Bennett,Peters},
etc.

Recently, the researching on MiN is arising: the monogamy of MiN is discussed \cite{Sen}; in \cite{Mirafzali}, a formula of
MiN for arbitrary (finite) dimensional system is proposed, the dynamics of MiN is explored in \cite{Humingliang}; MiN based on the relative entropy is discussed in \cite{Xizhengjun}; in \cite{GuoHou}, we extended it to infinite-dimensional case and the nullity of MiN is characterized, etc.
Xu in \cite{Xujianwei} introduced a geometric measure of quantum discord over two-sided projective measurements. In the same spirit, in this letter we introduce the two-sided MiN for both finite- and infinite-dimensional cases.

The paper is arranged as follows. In Sec.~II, we first introduce the origin MiN, namely, the MiN over one-sided projective measurements. Then introduc the geometric measure of quantum discord (GMQD) proposed in \cite{Dakic,GuoHou}. Sec.~III is devoted to establishing the MiN over two-sided projective measurements for both finite- and infinite-dimensional cases. Then in Sec.~IV we give a necessary and sufficient condition of a state has zero two-sided MiN. In Sec.~V, we introduce the two-sided GMQD for infinite-dimensional case and compare the two-sided MiN with that of GMQD.
A summarizing is given in the last section.

In this letter, we consider the two-mode system
labeled by A+B which is described by a
complex Hilbert space $H=H_A\otimes H_B$ with
$\dim H_A\otimes H_B\leq+\infty$.
We denote by $\mathcal{S}(H_A\otimes H_B)$ the
set of all states acting on $H_A\otimes H_B$.
\vspace{2mm}

\section{Measurement-induced nonlocality over one-sided projective measurements}

Before discussing the two-sided MiN after this section, we firstly review the origin one-sided
MiN in the present section.

MiN was firstly proposed
by Luo and Fu \cite{Luo1}, which can be viewed as a
kind of quantum correlation from a geometric perspective
based on the local projective measurements
from which one of the reduced states is left invariant.
The MiN of $\rho$ (with respect to part A), denoted by $N_A(\rho)$, is
defined by \cite{Luo1}
\begin{eqnarray}
N_A(\rho):=\max_{\Pi^a} \|\rho-\Pi^a(\rho)\|_2^2, \label{1}
\end{eqnarray}
where $\|\cdot\|_2$ stands for the Hilbert-Schmidt norm (that is
$\|A\|_2=[{\rm Tr}(A^\dag A)]^{\frac{1}{2}}$), and the maximum is
taken over all local projective measurement $\Pi^a=\{\Pi_k^a\}$
with $\sum_k\Pi_k^a\rho_A\Pi_k^a=\rho_A$,
where $\Pi^a(\rho)=\sum_k(\Pi_k^a\otimes I_B)\rho(\Pi_k^a\otimes I_B)$.

Recently, we extended MiN to infinite-dimensional case \cite{GuoHou} using the same scenario
as that of finite-dimensional case. That is,
\begin{eqnarray}
N_A(\rho):=\sup_{\Pi^a}\|\rho-\Pi^a(\rho)\|_2^2,
\end{eqnarray}
where the sup is taken over all local
projective measurement $\Pi^a=\{\Pi_k^a\}$
that satisfying
$
\sum_k\Pi_k^a\rho_A\Pi_k^a=\rho_A$.
Here, $\sum_k(\Pi_k^a\otimes I_B)^\dag(\Pi_k^a\otimes I_B)
=\sum_k\Pi_k^a\otimes I_B=I_{AB}$, and the series converges
under the strongly operator topology \cite{Hou}.

The MiN is in some sense dual to the geometric measure of quantum discord (GMQD) \cite{Luo1}.
GMQD is originally introduced in \cite{Dakic} as
\begin{eqnarray}
D_A^G(\rho):=\min_{\chi}\|\rho-\chi\|_2^2 \label{v}
\end{eqnarray}
with $\chi$ runs over all zero QD (up to part A) states.
Very recently, we extend GMQD to infinite-dimensional case in terms of the classical-quantum (CQ) states.
[Let $H_A$ and
$H_B$ are complex separable Hilbert spaces with $\dim H_A\otimes H_B\leq+\infty$.
Recall that
a state $\rho$ is called a CQ state if it can be written as \begin{eqnarray}
\rho=\sum_ip_i|i\rangle\langle i|\otimes\rho_i^B,
\end{eqnarray}
where $\{|i\rangle\}$ is an orthonormal
set of $H_A$, $\rho_i^B$s are
states of the subsystem B,
$p_i\geq0$ and $\sum_ip_i=1$ (for the finite-dimensional case, see in \cite{Luo4}).]
Assume that $\dim H_A\otimes H_B=+\infty$, we define the geometric
measure of quantum discord up to part A of a state by \cite{GuoHou}
\begin{eqnarray}
D_A^G(\rho)=\inf \{\|\rho -\pi\|_2^2 : \pi \in
{\mathcal CQ}\},
\end{eqnarray}
where ${\mathcal CQ}$ is the set of all CQ states on $H_A\otimes
H_B$. That is, the geometric quantum discord of a state $\rho$ is
the square of the Hilbert-Schmidt distance of the state  to the
set of all CQ states. $D_A^G(\rho)$ makes sense for any state
$\rho$ because states are Hilbert-Schmidt operators.
\vspace{2mm}

\section{Measurement-induced nonlocality over two-sided projective measurements}

We now define the MiN over two-sided projective measurements.
We call $\Pi^{ab}=\{\Pi^{\alpha\beta}\}$ a two-sided projective measurement if $\Pi^{\alpha\beta}=|\alpha\rangle\langle\alpha|\otimes|\beta\rangle\langle\beta|$,
where $\{|\alpha\rangle\}$ and $\{|\beta\rangle\}$ are orthonormal bases of $H_A$ and $H_B$ respectively, and where
$\sum_{\alpha,\beta}|\alpha\rangle\langle\alpha|\otimes|\beta\rangle\langle\beta|=I_A\otimes I_B$, the series converges under the strong operator topology \cite{Hou}.
Under the operation of $\Pi^{ab}$, $\rho$ becomes
\begin{eqnarray}
\Pi^{ab}(\rho)=\sum\limits_{\alpha,\beta}|\alpha\rangle\langle\alpha|\otimes|\beta\rangle\langle\beta|
\rho|\alpha\rangle\langle\alpha|\otimes|\beta\rangle\langle\beta|.
\end{eqnarray}
A natural way to define two-sided MiN then is
\begin{eqnarray}
N_{AB}(\rho):=\sup\limits_{\Pi^{ab}}\|\rho-\Pi^{ab}(\rho)\|_2^2, \label{2}
\end{eqnarray}
where the sup is taken over all two-sided projective measurements
$\{\Pi^{ab}\}$ that satisfying
\begin{eqnarray*}
\sum_\alpha|\alpha\rangle\langle\alpha|\rho_A|\alpha\rangle\langle\alpha|=\rho_A,~~
\sum_\beta|\beta\rangle\langle\beta|\rho_B|\beta\rangle\langle\beta|=\rho_B.
\end{eqnarray*}
If $\Pi^a=\{|\alpha\rangle\langle\alpha|\}$ is a local projective measurement on part A, and
$\Pi^b=\{|\beta\rangle\langle\beta|\}$ is a local projective measurement on part B, then
$\Pi^{ab}=\Pi^a\Pi^b=\Pi^b\Pi^a$ is a two-sided projective measurement and vice versa.

For the one-sided MiN, it is originally pointed out in \cite{Luo1} that,
\emph{it is particularly relevant to
certain cryptographic communication}.
The two-sided MiN may be also
useful in some quantum information tasks. For example, consider
the task that Alice and Bob are far away from each other and they want to send information to
each other. Assume that they share a joint state
$\rho$. Both Alice and Bob can encode their information by locally manipulating
her/his part of the state, and then sends it to each other, they
then decodes the message from the joint state.
If Charlie and Dave are eavesdroppers that snoop around Alice and Bob respectively.
In order to exclude eavesdropping in the communication processing, Alice and Bob choose measurements that will not disturb her/his
local state $\rho_{A/B}$.
Provided that $N_{AB}(\rho)\neq0$, consequently, Alice and Bob can
choose a measurement which maximizes the difference
between the pre- and post-measurement states in order
for both of them to detect the change of the joint state (thus
the encoded messages) most reliably. Charlie and Dave
can not get any information since Charlie is always facing the
same state $\rho_{A}$ and Dave is always facing the
same state $\rho_{B}$, however $\rho_{A}$ and $\rho_B$ are invariant.
In such a scenario, Alice and Bob can realize cryptographic communication
successfully.

According to the definition in Eq.~(\ref{2}), the following properties are straightforward.

(i) $N_{AB}(\rho)=0$ if and only if $N_A(\rho)=0$ and $N_B(\rho)=0$. Thus, $N_{AB}(\rho)=0$ for any product
state $\rho=\rho_A\otimes \rho_B$, and a pure state has nonzero
two-sided MiN if and only if it is entangled.

(ii) $N_{AB}(\rho)$ is locally unitary invariant, namely,
$N_{AB}[(U\otimes V)\rho(U^\dag\otimes V^\dag)]=N_{AB}(\rho)$
for any unitary operators $U$ and $V$
acting on $H_A$ and $H_B$, respectively.

(iii) $N_{AB}(\rho)>0$ whenever $\rho$ is
entangled since $\Pi^{ab}(\rho)$ is
always a classical state (the definition of classical state see in Sec.~V) and thus is separable.

(iv) $0\leq N_{AB}(\rho)<4$.

(v) If both $\rho_A$ and $\rho_B$ are nondegenerate, then the two-sided projective measurement that
make $\rho_{A/B}$ invariant is unique, and this unique one is just the one induced by the eigenvectors
of $\rho_A$ and $\rho_B$.

To evaluate the two-sided MiN is a hard work in general, but it is easy for pure states.\\

{\bf Theorem 1.}\quad Let $H_A$ and $H_B$ be complex separable Hilbert spaces with
$\dim H_A\otimes H_B\leq+\infty$, and let $|\psi\rangle\in H_A\otimes H_B$ be a pure state with the Schmidt decomposition
$|\psi\rangle=\sum_k\lambda_k|k\rangle|k'\rangle$. Then
\begin{eqnarray}
N_{AB}(|\psi\rangle)=N_A(|\psi\rangle)=N_B(|\psi\rangle)=1-\sum_k\lambda_k^4\leq1.
\end{eqnarray}\\

{\bf Proof.}\quad If $|\psi\rangle=\sum_k\lambda_k|k\rangle|k'\rangle$ is
the Schmidt decomposition of $|\psi\rangle$, then, for any local projective measurement
$\Pi^{ab}$ that leaves $\rho_A$ and $\rho_B$ invariant, one has
\begin{eqnarray*}
\begin{array}{rcl}
&&\Pi^{ab}(|\psi\rangle\langle\psi|)\\
&=&\sum\limits_{\lambda_k\neq\lambda_l}\lambda_k^2|k\rangle\langle k|\otimes |k'\rangle\langle k'|\\
&&+\sum\limits_{\lambda_k=\lambda_l}\lambda_k\lambda_l(\sum\limits_i\alpha_{ki}\bar{\alpha}_{li}|e_i\rangle\langle e_i|)\otimes(\sum\limits_j\beta_{kj}\bar{\beta}_{lj}|e_j'\rangle\langle e_j'|),
\end{array}
\end{eqnarray*}
where $\{|e_i\rangle\}$ (resp. $\{|e_j'\}$) satisfy
$\sum_i|e_i\rangle\langle e_i|=I_{E(\lambda_k)}$ (resp. $\sum_j|e_j'\rangle\langle e_j'|=I_{E'(\lambda_k)}$)whenever $\lambda_k=\lambda_l$, here $E(\lambda_k)$ (resp. $E'(\lambda_k)$) denotes the eigenspace of eigenvalue $\lambda_k^2$ of $\rho_A={\rm Tr}_B(|\psi\rangle\langle\psi|)$
(resp. $\rho_B={\rm Tr}_A(|\psi\rangle\langle\psi|)$), and where $\alpha_{ki}=\langle e_i|k\rangle$, $\beta_{lj}=\langle e_j'|l'\rangle$.
We compute
\begin{eqnarray*}
\begin{array}{rcl}
&&\||\psi\rangle\langle\psi|-\Pi^{ab}(|\psi\rangle\langle\psi|)\|_2^2\\
&=&\|\sum\limits_{k,l}\lambda_k\lambda_l|k\rangle\langle l|\otimes|k'\rangle\langle l'|-\sum\limits_{\lambda_k\neq\lambda_l}\lambda_k^2|k\rangle\langle k|\otimes |k'\rangle\langle k'|\\
&&-\sum\limits_{\lambda_k=\lambda_l}\lambda_k\lambda_l(\sum\limits_i\alpha_{ki}\bar{\alpha}_{li}|e_i\rangle\langle e_i|)\otimes(\sum\limits_j\beta_{kj}\bar{\beta}_{lj}|e_j'\rangle\langle e_j'|)\|_2^2\\
&=&{\rm Tr}[(\sum\limits_{k,l}\lambda_k\lambda_l|k\rangle\langle l|\otimes|k'\rangle\langle l'|-\sum\limits_{\lambda_k\neq\lambda_l}\lambda_k^2|k\rangle\langle k|\otimes |k'\rangle\langle k'|\\
&&-\sum\limits_{\lambda_k=\lambda_l}\lambda_k\lambda_l(\sum\limits_i\alpha_{ki}\bar{\alpha}_{li}|e_i\rangle\langle e_i|)\otimes(\sum\limits_j\beta_{kj}\bar{\beta}_{lj}|e_j'\rangle\langle e_j'|))^2]\\
&=&1+\sum\limits_k\lambda_k^4-2\sum\limits_{\lambda_k\neq\lambda_l}\lambda_k^4
-2\sum\limits_{\lambda_k=\lambda_l}\lambda_k^2\lambda_l^2\\
&=&1-\sum\limits_k\lambda_k^4.
\end{array}
\end{eqnarray*}
Together with the fact \cite{Luo1,GuoHou} $N_A(|\psi\rangle)=N_B(|\psi\rangle)=1-\sum\limits_k\lambda_k^4$, we complete the proof.
\hfill$\square$\\

Theorem 1 implies that $N_{AB}$ achieves the maximal value $\frac{m-1}{m}$ for maximally entangled pure state in $m\otimes n$ system with
$m\leq n$. \if Then, a question is: is there some mixed state $\rho$ such that $N_{AB}(\rho)$ exceed $\frac{m-1}{m}$ for $m\otimes n$ system with
$m\leq n$? We answer this question below.\fi Next, we consider the (two-sided) MiN for maximally entangled mixed states.
It is known that, for $m\otimes n$ ($m\leq n$)system, a pure state $|\psi\rangle$ is called maximally entangled
if $\rho_A=\frac{1}{m}I_A$. If a mixed state has the same entanglement quantified by a certain
entanglement measure as $|\psi\rangle$, we call it a maximally entangled mixed state \cite{Feishaoming}.\\

{\bf Theorem 2.}\quad Assume that $\dim H_A=m$, $\dim H_B=n$ and $2m\leq n$. If $\rho$ is a maximally entangled mixed state on $H_A\otimes H_B$, then
\begin{eqnarray}
N_{A}(\rho)=\frac{m-1}{m},~~N_{B}(\rho)\geq\frac{m-1}{m}, ~~N_{AB}(\rho)\geq\frac{m-1}{m}.\label{c}
\end{eqnarray} \\

{\bf Proof.}\quad
Let $\rho$ be a maximally entangled mixed state. By Theorem 1 in \cite{Feishaoming}, $\rho$ can be written as
\begin{eqnarray*}
\rho=\sum\limits_{k}p_k|\psi_k\rangle\langle\psi_k|,~~\sum\limits_kp_k=1,
\end{eqnarray*}
where $|\psi_k\rangle=\frac{1}{\sqrt{m}}\sum_{i=1}^m|i\rangle|i_k'\rangle$, $\{|i\rangle\}$ and $\{|i_k'\rangle\}$
are orthonormal bases of the subsystems A and B respectively, satisfying $\langle j_s|i_t\rangle=\delta_{ij}\delta_{st}$.
For simplicity, we write $\rho_A^{(i)}={\rm Tr}_B(|\psi_i\rangle\langle\psi_i|)$.
Obviously, $\rho_A=\rho_A^{(i)}=\frac{1}{m}I_A$ for any $i$, which reveals
that any local projective $\Pi^a$ make $\rho_A$ invariant, and thus make $\rho_A^{(i)}$ invariant.
Notice that for any local projective measurement $\Pi^a$, we have
\begin{eqnarray*}
{\rm Tr}(\Pi^{a}(|\psi_i\rangle\langle\psi_i|)\Pi^{a}(|\psi_j\rangle\langle\psi_j|))=0
\end{eqnarray*}
whenever $i\neq j$.
It follows that
\begin{eqnarray*}
\begin{array}{rl}
&\|\rho-\Pi^a(\rho)\|_2\\
=&\|\sum\limits_kp_k(|\psi_k\rangle\langle\psi_k|-\Pi^{a}(|\psi_k\rangle\langle\psi_k|))\|_2\\
=&\sum\limits_kp_k\||\psi_k\rangle\langle\psi_k|-\Pi^{a}(|\psi_k\rangle\langle\psi_k|)\|_2\\
=&\sum\limits_kp_k\sqrt{\frac{m-1}{m}}\\
=&\sqrt{\frac{m-1}{m}}
\end{array}
\end{eqnarray*}
holds for any local projective measurement $\Pi^a$,
that is $N_A(\rho)=\frac{m-1}{m}$ as desired.

Let $\rho_B^{(i)}={\rm Tr}_A(|\psi_i\rangle\langle\psi_i|)$.
For any local projective measurement $\Pi^b$ with $\Pi^b(\rho_B^{(i)})=\rho_B^{(i)}$,
we have
\begin{eqnarray*}
{\rm Tr}(\Pi^{b}(|\psi_i\rangle\langle\psi_i|)\Pi^{b}(|\psi_j\rangle\langle\psi_j|))=0
\end{eqnarray*}
whenever $i\neq j$
and $\Pi^b(\rho_B)=\rho_B$.
We assume further that $N_B(|\psi_k\rangle)=\||\psi_k\rangle\langle\psi_k|-\Pi^b(|\psi_k\rangle\langle\psi_k|)\|_2^2$.
Then
\begin{eqnarray*}
\begin{array}{rl}
&\|\rho-\Pi^b(\rho)\|_2\\
=&\|\sum\limits_kp_k(|\psi_k\rangle\langle\psi_k|-\Pi^{b}(|\psi_k\rangle\langle\psi_k|))\|_2\\
=&\sum\limits_kp_k\||\psi_k\rangle\langle\psi_k|-\Pi^{b}(|\psi_k\rangle\langle\psi_k|)\|_2\\
=&\sum\limits_kp_k\sqrt{\frac{m-1}{m}}\\
=&\sqrt{\frac{m-1}{m}},
\end{array}
\end{eqnarray*}
which leads to $N_B(\rho)\geq\frac{m-1}{m}$.

Replace $\Pi^b$ by $\Pi^{ab}$ and repeat the argument above, we can conclude
that $N_{AB}(\rho)\geq\frac{m-1}{m}$.
\hfill$\square$\\

{\bf Questions.}\quad Is it true if we replace `$\geq$' by `$=$' in Eq.~(\ref{c})?
Is $N_{A/B}$ (or $N_{AB}$) always smaller than $\frac{m-1}{m}$ for any $m\otimes n$ system with $m\leq n$?\\

By definition, it is obvious that $D_{A/B}^G$ and $D_{AB}^G$ are continuous functions on $\mathcal{S}(H_A\otimes H_B)$ with respect to trace norm (note that the trace norm topology coincides with the Hilbert-Schmidt norm topology on $\mathcal{S}(H_A\otimes H_B)$ \cite{Mazhihao}).
In the end of this section, we show that \\

{\bf Proposition 3.}\quad
$N_{A/B}$ and $N_{AB}$ are not continuous functions on
$\mathcal{S}(H_A\otimes H_B)$ with respect to the trace norm ($\dim H_A\otimes H_B\leq+\infty$). \\

We illustrate this fact with the example below. We consider the $m\otimes m$ Werner state
\begin{eqnarray}
\rho_x=\frac{m-x}{m^3-m}I_A\otimes I_B+\frac{mx-1}{m^3-m}F,\quad x\in[0,1], \label{p}
\end{eqnarray}
with $F=\sum_{i,j}|i\rangle\langle j|\otimes|j'\rangle\langle i'|$ is the flip operator.
Let $\sigma=(\sum_{i=1}^m\lambda_i|i\rangle\langle i|)\otimes(\sum_{j=1}^m\delta_j|j'\rangle\langle j'|)$ be a $m\otimes m$ state
with different $\lambda_i$s and $\delta_j$s such that the reduced sates $\sigma_A$ and $\sigma_B$ are nondegenerate.
In contrast, let $\varrho=(\sum_{i=1}^m\lambda_i|i\rangle\langle i|)\otimes(\sum_{j=1}^m\delta_j|\mu_j'\rangle\langle \mu_j'|)$ with $\{|\mu_j'\rangle\}$ is another orthonormal basis of $H_B$ such that $\{|j'\rangle\}$ and $\{|\mu_j'\rangle\}$ are
unbiased. \if (This example is borrowed from \cite{Wushengjun}, where it is used for demonstrating the classical correlation obtained by local projective measurements onto the eigenvectors of the reduced density matrices on both part A and part B is not continuous, see in \cite{Wushengjun} for detail).\fi
Let $\epsilon_n$ and $\varepsilon_n$ be infinitesimal numbers. Write
$\rho_n=\frac{1}{1+\epsilon}(\rho_x+\epsilon_n\sigma)$ and  $\varrho_n=\frac{1}{1+\epsilon_n}(\rho_x+\varepsilon_n\varrho)$.
Then the unique two-sided projective measurement $\Pi^{ab}$ on $\rho_n$ that leaves both $\rho_A^{(n)}$ and $\rho_B^{(n)}$ invariant is the one that induced from the eigenvector of $\rho_A^{(n)}$ and $\rho_B^{(n)}$, and the unique two-sided projective measurement $\Pi^{'ab} $ on $\varrho_n$ that leaves both $\varrho_A^{(n)}$ and $\varrho_B^{(n)}$ invariant is the one that induced from the eigenvector of $\varrho_A^{(n)}$ and $\varrho_B^{(n)}$
(here $\rho_{A/B}^{(n)}={\rm Tr}_{B/A}(\rho_n)$, $\varrho_{A/B}^{(n)}={\rm Tr}_{B/A}(\varrho_n)$. Note that $\rho_A^{(n)}$s (resp. $\varrho_A^{(n)}$s) have the same eigenvectors, and $\rho_B^{(n)}$s (resp. $\rho_B^{(n)}$s) have the same eigenvectors as well).
It is now clear that, although $\rho_n\rightarrow\rho_x$ and $\varrho_n\rightarrow\rho_x$ in the trace norm for some sequences $\{\epsilon_n\}$ and $\{\varepsilon_n\}$ (and thus $\varrho_n-\rho_n\rightarrow0$ in the trace norm),
$\|\varrho_n-\Pi^{'ab}(\varrho_n)\|_2-\|\rho_n-\Pi^{ab}(\rho_n)\|_2\nrightarrow0$, which implies that $N_{AB}(\varrho_n)-N_{AB}(\rho_n)\nrightarrow0$. Therefore, $N_{AB}$ is not continuous.
Similarly, one can check that $N_{A/B}$ is not continuous.
\vspace{2mm}

\section{Nullity of two-sided measurement-induced nonlocality}

Let $H_A$ and
$H_B$ be complex separable Hilbert spaces with $\dim H_A\otimes H_B\leq+\infty$, and let $\{|i\rangle\}$ and
$\{|i'\rangle\}$ be
the orthonormal bases of $H_A$ and $H_B$ respectively. Then any state $\rho$ acting on
$H_A\otimes H_B$ can be represented by
\begin{eqnarray}
\rho=\sum_{i,j}E_{ij}\otimes B_{ij}=\sum_{k,l}A_{kl}\otimes F_{kl},\label{z}
\end{eqnarray}
where $E_{ij}=|i\rangle\langle j|$, $F_{kl}=|k'\rangle\langle l'|$, $B_{ij}$s and $A_{kl}$s are trace class
operators on $H_B$ and $H_A$ respectively, and where the series converges in trace norm
\cite{Guo2}.

In \cite{GuoHou}, we proved that
\if $N_{A}(\rho)=0$ if and only if
\begin{eqnarray}
\rho=\sum_iq_i|i \rangle\langle i|\otimes\rho_i^B \label{y}
\end{eqnarray}
with $\rho_i^B=\rho_j^B$ whenever $p_j=p_i$ for some orthonormal basis $\{|i\rangle\}$ of $H_A$.
Symmetrically, $N_B(\rho)=0$ (resp. \fi
$N_A(\rho)=0$ if and only if $A_{kl}$s
\if (resp. $A_{kl}$)\fi
are mutually
commuting normal operators and
each eigenspace of $\rho_A$
\if (resp. $H_A$)\fi
contained in some eigenspace of
$B_{ij}$ for all $k$ and $l$.
\if(resp. $A_{kl}$ for all $k$ and $l$.\fi
Equivalently, we also obtain in \cite{GuoHou} that
$N_{A}(\rho)=0$ if and only if
\begin{eqnarray}
\rho=\sum_jp_j|j\rangle\langle j|\otimes \rho_j^B\label{x}
\end{eqnarray}
with $\rho_j^B=\rho_i^B$ whenever $p_j=p_i$, where $\{|j\rangle\}$ is an orthonormal basis of $H_A$.
Together with the fact $N_{AB}(\rho)=0$ if and only if $N_A(\rho)=N_B(\rho)=0$, the followings are obvious.\\

{\bf Theorem 4.}\quad Let $H_A$ and $H_B$ be complex separable Hilbert spaces with $\dim H_A\otimes H_B\leq+\infty$,
$\{|k\rangle\}$ and
$\{|i'\rangle\}$ be orthonormal bases
of $H_A$ and $H_B$, respectively,
and $\rho\in\mathcal{S}(H_A\otimes H_B)$.
Write $\rho=\sum_{i,j}E_{ij}\otimes B_{ij}=\sum_{k,l}A_{kl}\otimes F_{kl}$
as in Eq.~(\ref{z})
with respect to the given bases.
Then $N_{AB}(\rho)=0$ if and only if $B_{ij}$s and $A_{kl}s$ are mutually
commuting normal operators with the properties that
each eigenspace of $\rho_B$ (resp. $A_{kl}$) contained in some eigenspace of
$B_{ij}$ ($A_{kl}$) for all $i$ and $j$ (resp. $k$ and $l$).\\

Equivalently, we have\\

{\bf Theorem 5.}\quad Let $H_A$ and $H_B$ be complex separable Hilbert spaces with $\dim H_A\otimes H_B\leq+\infty$,
$\rho\in\mathcal{S}(H_A\otimes H_B)$. Then
$N_{AB}(\rho)=0$ if and only if
\begin{eqnarray}
\rho=\sum\limits_{i,j}p_{i,j}|i\rangle\langle i|\otimes|j'\rangle\langle j'|
\end{eqnarray}
with $p_{ij}=p_{ik}$ whenever $\sum_ip_{ij}=\sum_ip_{ik}$ and $p_{ij}=p_{lj}$ whenever $\sum_jp_{ij}=\sum_jp_{lj}$.\\

Assume that $\dim H_A\otimes H_B<+\infty$, $\dim H_A=m$ and $\dim H_B=n$. Let $\rho\in\mathcal{S}(H_A\otimes H_B)$ be a state
with $N_{AB}(\rho)=0$. Theorem 4 and 5 reveals that
\begin{eqnarray}
N_{AB}(t\rho+\frac{1-t}{mn}I_{AB})=0,\quad t\in[0,~1]
\end{eqnarray}
and
\begin{eqnarray}
N_{AB}(\rho^{T_{A/B}})=0,
\end{eqnarray}
where $\rho^{T_{A/B}}$ denotes the partial transpose of $\rho$.
One can also check that if $D_{AB}(\rho)=0$, then
\begin{eqnarray*}
D_{AB}(t\rho+\frac{1-t}{mn}I_{AB})=0,\quad t\in[0,~1]
\end{eqnarray*}
and
\begin{eqnarray*}
D_{AB}(\rho^{T_{A/B}})=0.
\end{eqnarray*}
Going further, (i) if $N_{A/B}(\rho)=0$, then
$
N_{A/B}(t\rho+\frac{1-t}{mn}I_{AB})=0$ for all $t\in[0,~1]$
and
$N_{A/B}(\rho^{T_{A/B}})=0$;
(ii) if $D_{A/B}(\rho)=0$, then
$D_{A/B}(t\rho+\frac{1-t}{mn}I_{AB})=0$ for all $t\in[0,~1]$
and
$D_{A/B}(\rho^{T_{A/B}})=0$.
That is, both the nullity of (two-sided) MiN and the nullity of (two-sided) QD are connected sets since all zero (two-sided) MiN states and all zero (two-sided) QD states are connected with the maximal mixed state
of the system. And the partial transpose of zero (two-sided) MiN (QD) state is still a zero (two-sided) MiN (QD) state.

\vspace{2mm}

\section{Comparing with (geometric measure of) quantum discord}

We now begin to discuss the relation between the two-sided MiN and the two-sided QD (GMQD).
For the finite-dimensional case, the QD over two-sided projective measurements is defined by \cite{Ollivier,Xujianwei}
\begin{eqnarray}
D_{AB}(\rho)=I(\rho)-\sup\limits_{\Pi^{ab}}I(\Pi^{ab}(\rho)),
\end{eqnarray}
where sup is taken over all two-sided projective measurements.
It is known that $D_{AB}(\rho)=0$ if and only if $\rho$ is a \emph{classical} state (see in \cite{Xujianwei} and
the references therein). Recall that, a bipartite state $\rho$ acting on $H_A\otimes H_B$ with $\dim H_A\otimes H_B\leq+\infty$ is called a
classical state, if it admits the form of
\begin{eqnarray}
\rho=\sum\limits_{i,j}p_{ij}|i\rangle\langle i|\otimes|j'\rangle\langle j'|,
\end{eqnarray}
where $\sum_{ij}p_{ij}=1$, $p_{ij}\geq0$.
For finite-dimensional system, the GMQD over two-sided projective measurements is defined by \cite{Xujianwei}
\begin{eqnarray}
D_{AB}^G(\rho)=\inf\limits_{\pi}\|\rho-\pi\|_2^2,
\end{eqnarray}
where inf is taken over all $\pi$ that $D_{AB}(\pi)=0$.
Let $\mathcal{C}$ be the set of all classical states in $\mathcal{S}(H_A\otimes H_B)$.
We now extend it to infinite-dimensional case by
\begin{eqnarray}
D_{AB}^G(\rho)=\inf\limits_{\varrho}\|\rho-\varrho\|_2^2,
\end{eqnarray}
where inf is taken over all $\varrho\in\mathcal{C}$. \if It can be derived from \cite{GuoHou} that
$\mathcal{C}$ in a closed set under the trace norm, thus $\varrho\in\mathcal{C}$ if and only if
$D_{AB}^G(\rho)=0$.
\fi

Denote by $\mathcal{N}_{AB}^0$, $\mathcal{D}_{AB}^0$ and $\mathcal{D}^{G-0}_{AB}$ the nullity of two-sided MiN, the nullity of two-sided QD and the nullity of two-sided GMQD, respectively.
We showed in \cite{GuoHou} that $D_A^G(\rho)=0$ if and only if $\rho$ is a CQ state for both finite- and infinite-dimensional cases.
Hence $\mathcal{D}_{AB}^0=\mathcal{D}_{AB}^{G-0}=\mathcal{C}$ holds for finite-dimensional case
and $\mathcal{D}_{AB}^{G-0}=\mathcal{C}$ holds for infinite-dimensional case.
Let $\mathcal{N}_A^0$ (resp. $\mathcal{N}_B^0$)be the set of all states with zero MiN with respect to part A (resp. B).
Recall that, a state $\rho\in\mathcal{S}(H_A\otimes H_B)$ is called a quantum-classical (QC) state,
if it can be expressed by $\rho=\sum_jq_j\rho_j^A\otimes|j'\rangle\langle j'|$
for some orthonormal basis $\{|j'\rangle\}$ of $H_B$, where $\{q_j\}$ is a probability distribution, $\rho_j^A$s are states on $H_A$.
Let  $\mathcal{QC}$ be the set of all QC states, then $\mathcal{C}=\mathcal{CQ}\cap\mathcal{QC}$.
In \cite{GuoHou}, we proved that $\mathcal{N}_A^0$ (resp.  $\mathcal{N}_B^0$) is a proper subset of $\mathcal{CQ}$ (resp. $\mathcal{QC}$).
Together with Theorem 5, we can thus get that\\

{\bf Theorem 6.} \quad \if Let $H_A$ and $H_B$ be complex separable Hilbert spaces with
$\dim H_A\otimes H_B\leq+\infty$.\fi
$\mathcal{N}_{AB}^0$ is a proper subset of $\mathcal{D}_{AB}^{G-0}$
for both finite- and infinite-dimensional cases.\\

For finite-dimensional case, analytic formulas for calculating the GMQD and the two-side GMQD are proposed \cite{Luoshun,Xujianwei}. Let $\rho\in\mathcal{S}(H_A\otimes H_B)$ with $\dim H_A\otimes H_B<+\infty$. Assume that $\dim H_A=m$ and $\dim H_B=n$. We denote by $\mathcal{B}(H_{A/B})$ the space of all liner operators on $H_{A/B}$. For given sets of Hermitian operators
$\{X_i:~i=1,2,\dots,m^2\}$ and $\{Y_j:~j=1,2,\dots,n^2\}$ that constitute orthonormal
Hilbert-Schmidt bases of $\mathcal{B}(H_A)$ and $\mathcal{B}(H_B)$ respectively, $\rho$ can be decomposed as
$\rho=c_{ij}X_i\otimes Y_j$. For any orthonormal bases $\{|k\rangle\}$ of $H_A$
and $\{|l'\rangle\}$ of $H_B$, let $|k\rangle\langle k|=\sum\limits_{i=1}^{m^2}a_{ki}X_i$,
$|l'\rangle\langle l'|=\sum\limits_{j=1}^{n^2}b_{lj}Y_j$, $A=[a_{ki}]$, $B=[b_{lj}]$ and $C=[c_{ij}]$.
It is showed in \cite{Luoshun} that
\begin{eqnarray*}
&D_{A}^G(\rho)={\rm Tr}(CC^t)-\sup\limits_{A,B}{\rm Tr}(ACC^tA^t),&\\
&D_{A}^G(\rho)\geq{\rm Tr}(CC^t)-\sum\limits_{k=1}^{m}\lambda_k.&
\end{eqnarray*}
In \cite{Xujianwei}, the $D_{AB}^G(\rho)$ is evaluated by
\begin{eqnarray*}
&D_{AB}^G(\rho)={\rm Tr}(CC^t)-\sup\limits_{A,B}{\rm Tr}(ACB^tBC^tA^t),&\\
&D_{AB}^G(\rho)\geq{\rm Tr}(CC^t)-\sum\limits_{k=1}^{\min\{m,n\}}\lambda_k,&\label{b}
\end{eqnarray*}
where $\lambda_k$ are eigenvalues of $CC^t$ listed in decreasing order (counting multiplicity),
$t$ denotes transpose.

If $\rho_A=\frac{1}{m}I_A$, then all local projective measurements $\Pi^a$ leave $\rho_A$ invariant.
In such a case, it is immediate from Eqs.~(\ref{1}) and (\ref{v}) that $N_A(\rho)\geq D_A^G(\rho)$.
Symmetrically, if  $\rho_B=\frac{1}{n}I_B$, one has $N_B(\rho)\geq D_B^G(\rho)$.
And of course that, $N_{AB}(\rho)\geq D_{AB}^G(\rho)$ provided that $\rho_A=\frac{1}{m}I_A$ and $\rho_B=\frac{1}{n}I_B$. The following result is easy to check.\\

{\bf Theorem 7.}\quad Let $\rho\in\mathcal{S}(H_A\otimes H_B)$ be a state with
$\dim H_A=m<+\infty$ and $\dim H_B=n<+\infty$. Then the following statements are true:

(i) If $\rho_A=\frac{1}{m}I_A$,
then
\begin{eqnarray*}
&N_{A}(\rho)={\rm Tr}(CC^t)-\inf\limits_{A,B}{\rm Tr}(ACC^tA^t)\geq D_A(\rho),&\\
&N_{A}(\rho)\geq D_A(\rho)\geq{\rm Tr}(CC^t)-\sum\limits_{k=1}^{m}\lambda_k;&
\end{eqnarray*}

(ii) If $\rho_B=\frac{1}{n}I_B$,
then
\begin{eqnarray*}
&N_{B}(\rho)={\rm Tr}(CC^t)-\inf\limits_{A,B}{\rm Tr}(CB^tBC^t)\geq D_B(\rho),&\\
&N_{B}(\rho)\geq D_B(\rho)\geq{\rm Tr}(CC^t)-\sum\limits_{k=1}^{n}\lambda_k;&
\end{eqnarray*}

(iii) If $\rho_A=\frac{1}{m}I_A$ and $\rho_B=\frac{1}{n}I_B$, then
\begin{eqnarray*}
&N_{AB}(\rho)={\rm Tr}(CC^t)-\inf\limits_{A,B}{\rm Tr}(ACB^tBC^tA^t)\geq D_{AB}(\rho),&\\
&N_{AB}(\rho)\geq D_{AB}(\rho)\geq{\rm Tr}(CC^t)-\sum\limits_{k=1}^{\min\{m,n\}}\lambda_k.&
\end{eqnarray*}
\vspace{2mm}

\section{Examples}

At last, we present some examples to illustrate the nullity of two-sided MiN and
compare it with the two-sided QD(GMQD).\\

{\bf Example 1.}\quad We consider the two-qubit Bell-diagonal states
\begin{eqnarray*}
\rho=\frac{1}{4}(I_A\otimes I_B+\sum\limits_{i=1}^{3}\sigma_i\otimes\sigma_i),
\end{eqnarray*}
where $c_i$s are real numbers, $\sigma_i$s are Pauli matrices, i.e., $\sigma_1=\left(\begin{array}{cc}0&1\\1&0\end{array}\right)$,
$\sigma_2=\left(\begin{array}{cc}0&-i\\i&0\end{array}\right)$ and $\sigma_3=\left(\begin{array}{cc}1&0\\0&-1\end{array}\right)$.
It is showed in \cite{Xujianwei} that $D_{AB}^G(\rho)=0$ if and only if $c_1=c_2=0$.
However, one can check that $N_{AB}(\rho)=0$ if and only if $c_1=c_2=c_3=0$.
Also note that $\rho_A=\rho_B=\frac{1}{2}\left(\begin{array}{cc}1&0\\0&1\end{array}\right)$.
That is $D_{AB}^G(\rho)\geq N_{AB}(\rho)$ for Bell-diagonal state $\rho$ and there exit Bell-diagonal
states such that they have quantum correlation tested by two-sided QD while they don't contain quantum correlation
quantified by two-sided MiN.\\

However, as what we will show,
for the so-called Werner states and the isotropic states, the zero two-sided MiN coincides
with the zero two-sided QD(GMQD).\\

{\bf Example 2.} \quad For the $m\otimes m$ Werner state as in Eq.~(\ref{p}),
it is showed that
$D_A^{G}(\rho_x)=0$ if and only if $x=\frac{1}{m}$ \cite{Luoshun}, $D_{AB}^{G}(\rho_x)=0$ if and only if $x=\frac{1}{m}$ \cite{Xujianwei}.
Clearly, $N_{AB}(\rho_x)=0$ if and only if $x=\frac{1}{m}$.
Namely, the nullity of (two-sided) MiN coincides with that of (two-sided) QD(GMQD), which is the trivial one, i.e.,
$\rho$ is the maximally mixed state.\\

{\bf Example 3.}\quad For the $m\otimes m$ isotropic state
\begin{eqnarray*}
\rho_x=\frac{1-x}{m^2-1}I_A\otimes I_B+\frac{m^2x-1}{m^2-1}|\psi^+\rangle\langle\psi^+|,\quad x\in[0,1],
\end{eqnarray*}
with $|\psi^+\rangle$ is the maximally entangled state. It is known that
$D_A^{G}(\rho_x)=0$ if and only if $x=\frac{1}{m^2}$ \cite{Luoshun} and $D_{AB}^{G}(\rho_x)=0$ if and only if $x=\frac{1}{m^2}$ \cite{Xujianwei}.
Apparently, $N_{AB}(\rho_x)=0$ if and only if $x=\frac{1}{m^2}$.
So the nullity of (two-sided) MiN coincides with that of (two-sided) QD(GMQD).
\vspace{2mm}

\section{Summary}

With the same sprit as the two-sided geometric measure of
quantum discord introduced in \cite{Xujianwei}, and in some sense dual to it,
the two-sided measurement-induced nonlocality is established for both finite- and infinite-dimensional cases.
Consequently, an analytic formula of $N_{AB}$ for
pure states is given, the nullity of the two-sided
MiN is explored, and show that
there exist nonzero two-sided MiN states that don't have two-sided QD.
Furthermore, the two-sided GMQD for infinite-dimensional system is proposed, and then we compare it with (two-sided) MiN. We thus obtain a clear structure of these two different quantum correlations.

Generally, entanglement measure is continuous, for example, concurrence is such a situation \cite{Guo3}, and the Negativity \cite{Vidal} is continuous as well.
However, in contrast to it, we illustrate with example that both the origin one-sided MiN and the two-sided MiN
are not continuous functions of the states.

Except for the questions proposed in Sec.~III, there are some other issues that are worth further discussion.
The first one is to comparing the dynamics of (two-sided) MiN with that of (two-sided) QD(GMQD).
The second concerns the comparing of MiN with the well-accepted entanglement measure, entanglement of formation \cite{Wootters} or concurrence \cite{Guo3,Chen2}. These works will provide us a more explicit picture of these different quantum correlations.\\

{\bf Acknowledgements.}\  This work is partially supported by
Natural Science Foundation of China (11171249,11101250) and Research
start-up fund for the Doctor of Shanxi Datong University.
\if The authors thanks Pro. C.-K. Li for helpful discussions.\fi

\end{document}